\documentclass{elsart2}
\usepackage{epsfig}
\usepackage{amssymb}

\def\perkeo{PERKEO\nolinebreak\hspace*{.3ex}II }

\begin{document}
\begin{frontmatter}
\title{Neutron $\beta$-decay
and the quark-mixing CKM-matrix}
\author{Hartmut Abele}
\address {Physikalisches Institut der Universit\"at Heidelberg,\\
Philosophenweg 12, D-69120 Heidelberg, Germany}
\begin{abstract}
Neutrons study a number of hot topics from the field of particle
physics and cosmology: Recent measurements of various mixed
American-British-French-German-Hungarian-Japan-Russian groups of
researchers determine the strength of the weak interaction of the
neutron, which gives us unique information on quark-mixing and the
question of unitarity. Much to our surprise, with neutron-decay we
find a deviation $\Delta$ = 0.0083(28) from the unitarity
condition, an effect that cannot be explained by the current
Standard Model of particle physics.
\end{abstract}
\end{frontmatter}
\section{Up, charm, top and quark mixing}
According to the accepted theory of particles and fields, matter
is built from two types of fundamental particles, called quarks
and leptons. The strong interaction glues quarks together to
hadrons. These quarks are considered to be quantum mechanical mass
eigenstates. When such a hadron decays due to the electroweak
interaction, the quarks being involved in the process of weak
interaction do mix and the mixing is expressed in the so-called
CKM-matrix. By convention, the u, c and t quarks are unmixed and
all mixing is expressed via the CKM-matrix operating on d, s and b
quarks. The Standard Model of elementary particle physics requests
that the mixing ends up in a zero-sum, in other words, the
CKM-quark-mixing matrix has to be unitary. Unitarity requires that
the sum of the squares of the matrix elements for each row and
column be unity.

We have studied the mixing of the down quark in the decay of free
neutrons. Much to our surprise, with new neutron-decay data the
zero-sum of quark-mixing ends up with a significant deviation
$\Delta$ = 0.0083(28), which is three times the stated error, an
effect that cannot be explained by the current Standard Model of
particle physics \cite{Abele1}.
\section{Neutron $\beta$-decay}
In the Standard Model, two free parameters describe neutron
$\beta$-decay. One parameter is the already mentioned first entry
$|V_{ud}|$ of the CKM-matrix. The other one is $\lambda$, the
ratio of the vector coupling constant and the axial vector
constant. In neutron decay, several observables are accessible to
experiment, which depend on these parameters, so the problem is
overdetermined and, together with other data from particle and
nuclear physics, many tests of the Standard Model become possible.
The chosen observables for determining $|V_{ud}|$ are the neutron
lifetime $\tau$ and a measurement of the $\beta$-asymmetry
parameter $A_0$. The lifetime is given by
\begin{equation}
\tau^{-1}=C |V_{ud}|^2(1+3\lambda^2) f^R(1+\Delta_R),
\end{equation}
where $C=G_F^2 m_e^5/(2 \pi^3)=1.1613\cdot10^{-4} s^{-1}$ in
$\hbar=c=1$ units, $f^R$ =1.71482(15) is the phase space factor
\cite{Wilkinson1} (including the model independent radiative
correction) adjusted for the current value of the neutron-proton
transition energy. $\Delta_R$ = 0.0240(8) is the model dependent
radiative correction to the neutron decay rate. The
$\beta$-asymmetry $A_0$ is linked to the probability that an
electron is emitted with angle $\vartheta$ with respect to the
neutron spin polarization $P$ = $<\sigma_z>$: \cite{Jackson}
\begin{equation} W(\vartheta) = 1 +\frac{v}{c}PA\cos(\vartheta),
\end{equation} where $v/c$ is the electron velocity expressed in
fractions of the speed of light. ${\it A}$ is the
$\beta$-asymmetry coefficient which depends on $\lambda$. On
account of order 1\% corrections for weak magnetism, $g_V-g_A$
interference,
and nucleon recoil, ${\it A}$ has the form $A$ = $A_0$(1+$A_{{\mu}m}$($A_1W_0+A_2W+A_3/W$)) 
with electron total energy $W = E_e /m_ec^2+1$ (endpoint $W_0$).
$A_0$ is a function of $\lambda$
\begin{equation}  A_0=-2\frac{\lambda(\lambda+1)}{1+3\lambda^2},
\end{equation} where we have assumed that $\lambda$ is real. The coefficients $A_{{\mu}m}$, $A_1$, $A_2$, $A_3$ are from \cite{Wilkinson1} taking a different $\lambda$ convention into consideration. In addition, a further small radiative
correction \cite{Gluck} of order 0.1\% must be applied.
\section{The experiment PERKEO and the result for $|V_{ud}|$}
The strategy of \perkeo followed the instrument PERKEO~\cite{Bopp}
in minimizing background and maximizing signal with a $4\pi$ solid
angle acceptance over a large region of the beam. Major
achievements of the instrument \perkeo are:
\begin{itemize}
\item The signal to background ratio in the range of interest is
200. \item The overall correction of the raw data is 2.04\%. \item
The detector design allows an energy calibration with linearity
better than 1\%. \item New polarizers and developments in
polarization analysis led to smaller uncertainties related to
neutron beam polarization.
\end{itemize}
\begin{figure}[tbh]
\noindent \centering \epsfig{file=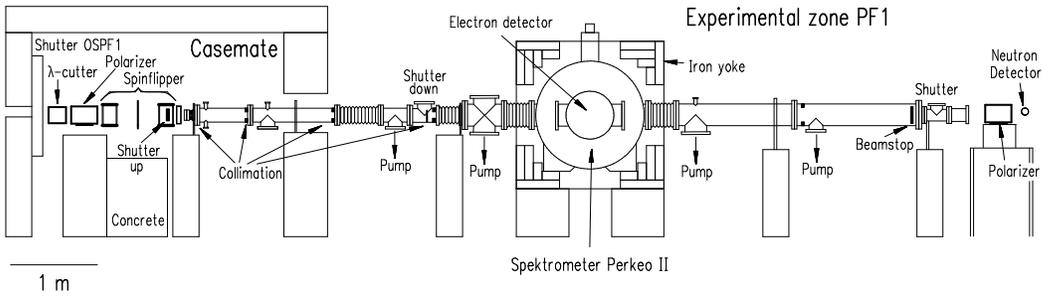, width=\linewidth, clip=} %
\caption{\small A schematic view of the whole setup at the ILL.}
\label{st97}
\end{figure}
For a measurement of $\beta$-asymmetry $A_0$, the instrument
PERKEO was installed at the PF1 cold neutron beam position at the
High Flux Reactor at the Institut Laue-Langevin, Grenoble. Cold
neutrons are obtained from a 25 K deuterium cold moderator near
the core of the 57 MW uranium reactor. The neutrons are guided via
a 60 m long neutron guide of cross section 6 x 12 cm$^2$ to the
experiment and are polarized by a 3 x 4.5 cm$^2$ supermirror
polarizer. The de Broglie wavelength spectrum of the cold neutron
beam ranges from about 0.2 nm to 1.3 nm. The degree of neutron
polarization was measured to be P = 98.9(3)\% over the full cross
section of the beam. The polarization efficiency remained constant
during the whole experiment. The neutron polarization is reversed
periodically with a current sheet spin flipper. The main component
of the PERKEO II spectrometer is a superconducting 1.1 T magnet in
a split pair configuration, with a coil diameter of about one
meter. Neutrons pass through the spectrometer, whereas decay
electrons are guided by the magnetic field to either one of two
scintillation detectors with photomultiplier readout. The detector
solid angle of acceptance is truly 2x2$\pi$ above a threshold of
60 keV. Electron backscattering effects, serious sources of
systematic error in $\beta$-spectroscopy, are effectively
suppressed. Technical details about the instrument can be found in
\cite{Abele}. The measured electron spectra $N^\uparrow_i(E_e)$
and $N^\downarrow_i(E_e)$ in the two detectors (i=1,2) for neutron
spin up and down, respectively, define the experimental asymmetry
as a function of electron kinetic energy $E_e$ and are shown in
Fig. 2. \begin{equation} A_{i_{exp}}(E_e)=\frac{N^\uparrow_i(E_e)
- N^\downarrow_i(E_e)}{N^\uparrow_i(E_e) + N^\downarrow_i(E_e)}.
\end{equation}
By using (4) and with $<\cos(\vartheta)>$ = 1/2, $A_i{_{exp}}(E)$
is directly related to the asymmetry parameter
\begin{equation}
A_{exp}(E_e)=A_{1_{exp}}(E_e)-A_{2_{exp}}(E_e)=\frac{v}{c}APf.
\end{equation}
\begin{figure}[h]
\noindent \centering \epsfig{figure=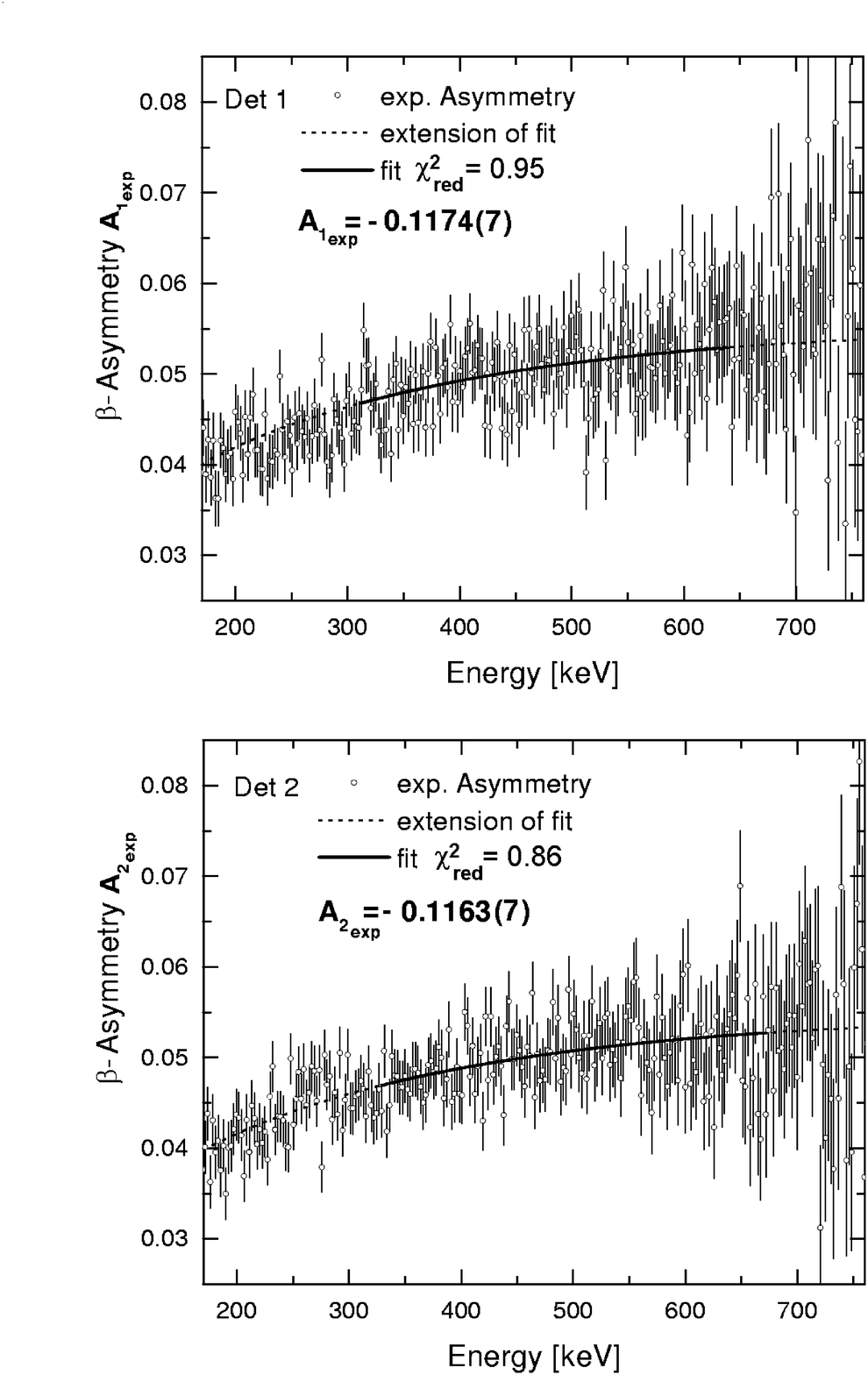,
width=6cm}\caption{\small Fit to the experimental asymmetry
$A_{exp}$ for detector 1 and detector 2. The solid line shows the
fit interval, whereas the dotted line shows an extrapolation to
higher and lower energies.} \label{fit}
\end{figure}
The experimental function $A_{i_{exp}}(E_e)$ and a fit with one
free parameter $A_{i_{exp}}$ (the absolute scale of $A_0$) is
shown in Fig. \ref{fit}. The total correction for the small
experimental systematic effects is 2.04\%.

With recent experiments from the University of Heidelberg
\cite{Abele1,Abele}, we obtain $A_0$ = -0.1189(7) and $\lambda$ =
- 1.2739(19). With this value, and the world average for $\tau$ =
885.7(7) s, we find that $|V_{ud}|$ = 0.9713(13). With $|V_{us}|$
= 0.2196(23) and the negligibly small $|V_{ub}|$ = 0.0036(9), one
gets
\begin{equation} |V_{ud}|^2 +
|V_{us}|^2 + |V_{ub}|^2 = 1 - \Delta = 0.9917(28).
\end{equation} This value differs from the Standard
Model prediction by $\Delta$ = 0.0083(28), or three times the
stated error. Earlier experiments
\cite{Yerozolimsky,Schreckenbach,Bopp} gave significant lower
values for $\lambda$. Averaging over our new result and previous
results, the Particle Data Group \cite{Groom} arrives at a new
world average for $|V_{ud}|$ from neutron $\beta$-decay which
leads to a 2.2 s deviation from unitarity. Usually, $|V_{ud}|$ is
derived from superallowed nuclear $\beta$-decay experiments and
this value of $|V_{ud}|$ includes nuclear structure effect
corrections. Combined with kaon, hyperon- and B-decays, this leads
to $\Delta$ =0.0032(14), signaling a deviation from the unitarity
condition by 2.3 standard deviations \cite{Hardy}. An independent
test of CKM unitarity comes from W physics at LEP \cite{Sbarra}
where W decay hadronic branching ratios can be used expressed in
terms of
\begin{equation}
\frac{Br(W\rightarrow q\bar{q})}{1-Br(W\rightarrow
\bar{q})}=(1+\frac{\alpha}{\pi}\sum{|V_{ij}|^2}).
\end{equation}
Since decay into the top quark channel is forbidden by energy
conservation one would expect $\sum{|V_{ij}|^2}$ to be 2 with a
three generation unitary CKM matrix. The experimental result is
2.032(32), consistent with (6) but with considerably lower
accuracy.
\section{Correlation $B$ and a search for right handed currents}
Parity is maximally violated in low energy physics. However, we do
not have a fundamental justification for parity violation. It is
particularly interesting that modern grand-unified theories
support a left-right symmetrical universe right after the start of
the big bang. Parity violation arises only due to a spontaneous
symmetry breaking at some intermediate energy scale. Parity
violation is not 100\% and right handed contributions in the weak
interaction should be found. Measurements of the correlation
coefficient $B$, the correlation between neutrino momentum and
neutron spin, are sensitive to right handed current contributions
in the weak interaction. However, we have no evidence for right
handed currents so far. The spectrometer PERKEO II has been
installed at the new beam position PF1B for a measurement of
coefficient B. The basic principle of a coefficient B measurement
is to measure the charged decay particles in neutron decay in
order to reconstruct the neutrino momentum with respect to the
neutron spin. Usually this is done with one electron and one
proton detector. PERKEO has one electron detector and one proton
detector in each hemisphere. This is an advantage over other
experiments because it maximizes the sensitivity on B. What is
more, B shows in a reasonable region no energy dependence on the
decay electrons. Systematic errors due to the detector response
function are small. The proton is measured in coincidence with a
decay electron. The $\beta$-detectors are made of plastic
scintillators. The proton detectors also make use of the
$\beta$-detectors. The idea is to convert a proton into an
electron signal. A proton will be accelerated up to 30 keV and
eventually hit a thin foil of carbon. About five secondary
electrons will be created and detected with the electron
detectors. This method of proton detection was already used by
Stratowa et al. \cite{Stratowa}. for a measurement of coefficient
a. Measurements are underway and first results are expected soon.
\section{Summary}
In summary, $|V_{ud}|$, the first element of the CKM matrix, has
been derived from neutron decay experiments in such a way that an
unitarity test of the CKM matrix can be performed based solely on
particle physics data. With this value, we find a 3 $\sigma$
standard deviation from unitarity, which conflicts the prediction
of the Standard Model of particle physics. This work was funded by
the German Federal Ministry for Research and Education under
contract number 06HD953.

\vfill


\begin{thebibliography}{00}
\bibitem{Abele1}H. Abele et al., Phys. Rev. Lett. 88 (2002)
211801.
\bibitem{Wilkinson1}D.H. Wilkinson, Nucl. Phys. A 377 (1982) 474; S. M. Bilenkii et al., Soviet. Phys. JETP 37 (1960) 1241.
\bibitem{Jackson}J.D. Jackson et al., Phys. Rev. 106 (1957) 517.
\bibitem{Gluck}F. Gl\"{u}ck and K. To'th, Phys. Rev. D 46 (1992) 2090; R. T. Shann, Nuovo Cimento A5 (1971) 591.
\bibitem{Bopp}P. Bopp et al., Phys. Rev. Lett. 56 (1988) 919.
\bibitem{Abele}H. Abele et al., Phys. Lett. B 407 (1997) 212.
\bibitem{Yerozolimsky}B.G. Yerozolimsky et al. Phys. Lett. B 412 (1997)
240; B.G. Erozolimskii et al., Phys. Lett. B 263 (1991) 33.
\bibitem{Schreckenbach}K. Schreckenbach et al., Phys. Lett. B 349 (1995) 427; P. Liaud et al., Nucl. Phys. A 612 (1997) 53.
\bibitem{Groom}D.E. Groom et al., (Particle Data Group), Eur. Phys. J. C 15 (2000) 1 and 2001 off-year partial update for the 2002 edition available on the PDG WWW pages.
\bibitem{Hardy}J. Hardy et al., nucl-th/9812036 (1998).
\bibitem{Sbarra}C.Sbarra, talk at the Rencontre de Moriond, Les Arcs, Savoie, France (March 11-18, 2000).
\bibitem{Stratowa}C. R. Stratowa et al., Phys. Rev. D 18 (1978)
3970.






\end{thebibliography}
\end{document}